\documentclass[preprint,12pt,showpacs]{revtex4} 
\usepackage{amsmath}
\usepackage{latexsym}
\usepackage{amssymb}
\usepackage{graphics,epstopdf}
\usepackage{amsmath,amssymb,amsthm}
\usepackage{slashed}
\usepackage[colorlinks=true, citecolor=blue, urlcolor=blue ]{hyperref}
\begin{document}
\title{Entropy uncertainty principle for Dirac system with mass jump }
\author{Pinaki Patra}
\email{monk.ju@gmail.com}
\affiliation{Department of Physics, Brahmananda Keshab Chandra College, Kolkata, India-700108}
\author{Kalpana Biswas}
\thanks{Corresponding author}
\email{klpnbsws@gmail.com}
\affiliation{Department of Physics, University of Kalyani, West Bengal, India-741235}
\affiliation{Department of Physics, Sree Chaitanya College, Habra, North 24 Parganas, West Bengal-743268}

\date{\today}

\begin{abstract}
Dependency on the preparation of state for the Heisenberg uncertainty principle can be removed with the help of entropy uncertainty principle. The shortness of the uncertainty principle (UP) can be overcome with the help of the concept of Shannon's information entropy (SE).  \\
In this article, we have shown that UP in terms of SE holds for a position-dependent effective mass system.  We have considered the Dirac system with a mass-jump at the origin. We have proved the existence of a lower bound for a UP for this position-dependent effective mass.
\end{abstract}

 \maketitle
 \section{Introduction}
   Heisenberg's uncertainty principle (HUP) states the limitations of one's ability to perform measurements on a system without disturbing it\cite{Heisenberg1927,Folland,Busch2009,Alladi,Beckner,HUP1}. According to HUP, there exist incompatible observables, simultaneous or sequential measurement outcome of which must have some degree of unsharpness. The inability of simultaneous sharp measurements of incompatible observables is not due to the limitation of present-day technology. It is a restriction of nature put upon us \cite{Busch, BuschP, Trina, Carruthers, Louisell}. For the incompatible observables which are connected by Fourier transformation (FT), the HUP is an immediate consequence of Paley -Weigner theorem (PW).  According to PW, FT of a compactly supported distribution will not be compactly supported \cite{Beckner,Alladi,Busch,BuschP}. For example, position and momentum observables in Quantum mechanics (QM), are connected by FT. If one takes the measure of sharpness of measurement outcome of observables in terms of standard deviation, then the mathematical expression of position $(x)$ and momentum $(p)$ uncertainty relation can be written as 
  \begin{equation}\label{xpuncert}
  \Delta x \Delta p \ge \frac{\hbar}{2}.
   \end{equation}
   In general, for two observables $A$ and $B$, the mathematical form of HUP can be written in terms of expectation value $(\langle\mathcal{O}\rangle)$ of the commutator of the observables $(\mathcal{O})$. In particular
    \begin{equation}\label{ABuncert}
    \Delta A \Delta B \ge \frac{1}{2}\vert \langle \left[A,B\right] \rangle \vert.
    \end{equation}
     Drawback of this mathematical relation is that it depends on the state preparation \cite{Busch,BuschP,Trina,Carruthers,Louisell,Deutsch,Drawback1}. For example if we prepare a system in a eigenstate of the spin-$\frac{1}{2}$ measurement observable in $z$-direction ($S_z=\frac{1}{2}\sigma_z$), then $\Delta S_z =0$, hence $\Delta S_z \Delta S_y =0$, which defies the fact of noncommutativity of Pauli matrices $\sigma_z$ and $\sigma_y$. These paradoxial situations can be overcome with the help of a stronger version of HUP. One of the stronger version of uncertainty principle\cite{Bialynicki}  was fulfilled by using Shannon information entropy (SE)\cite{Shannon} 
     \begin{equation}\label{shannondef0}
     S=-\sum_{i}p_i\log p_i,
     \end{equation}
    where $p_i $  being the probability of $i^{\mbox{th}}$ outcome. Later, the concept of \cite{Bialynicki} was used extensively in the literature \cite{Kraus,Maasen,Wehner}. Using the probability density $\rho=\psi^*\psi =\vert\psi\vert^2$, one can write the version of SE for QM in position $(x)$ reprentation as 
  \begin{equation}\label{shannondefn}
  S_x=-\langle \log\rho\rangle=-\int \vert\psi\vert^2\log\vert\psi\vert^2 dx .
  \end{equation}
  Similarly, one can write SE for QM in momentum $(p)$ reprentation as
  \begin{equation}\label{shannondefnp}
   S_p=-\langle \log \tilde{\rho}\rangle=-\int \vert\phi\vert^2\log\vert\phi\vert^2 dp ,
   \end{equation}
   where $\phi(p)$ is the Fourier transformation of $\psi(x)$. \\
  In terms of the entropies $S_x$ and $S_p$ in $n$-dimensional position and momentum space respectively, one of the stronger version of the uncertainty relation for wave mechanics is given by \cite{Bialynicki}
  \begin{equation}\label{uncertaintydefinition}
 S_x+S_p\ge n(1+\log\pi) .
  \end{equation}
 Since the uncertainty measure is additive in \eqref{uncertaintydefinition}, it is nonzero even if one term is zero; thus no ambiguity occurs.. \\
 The constructions of SE for various Quantum mechanical systems are available in the literature. However, the study of this regime was confined mainly to the constant mass case. The study of SE uncertainty relation for position-dependent effective mass was mainly confined in the examples related to the toy models where the knowledge of solution was known by Supersymmetric QM (SUSY) or by Quantum canonical transformations (QCT) \cite{qinfo1,qinfo2,qinfo3,qinfo4,qinfo5,qinfo6,Pinaki3}.  However, these toy models of mass distribution are mostly chosen for mathematical simplicity so that the SUSY or QCT may be applied to find close form solutions. The real-life examples suitable for the experimental purpose is lacking. In particular, the case of a mass profile which has a discontinuous jump at some point has not been studied in the shed of SE. This article aims to fulfill this gap. In this article, we have verified that the entropic uncertainty relation holds for the Dirac System with a mass jump. In the first section, a brief description of the systems with position-dependent effective mass is given. Then the entropic uncertainty relation is tallied for the particular system.
 \section{position dependent effective mass}
 Position dependent effective mass (PDEM) was first considered in the description of electronic properties and band structure of semiconductor Physics \cite{Pinaki,CostaFilho,Muharimousavi,Souza Dutra,Souza Dutra1,Schmidt,Abdalla,Jha,Jha1,heterostructure1,heterostructure2,Mario,yu,lozada,arias,guedes,mello,santos,cavalcanti,cunha,bekke,vitoria,vitoria1,bekke1,delta1,optical properties,optical properties2,crackling noise,crackling2,qinfo1,qinfo2,qinfo3,qinfo4,qinfo5,qinfo6}. Later it was proved to be an useful tool in other branches of Physics and growing interest in this regime had made the concept of PDEM a tropical one \cite{tdm1,tdm2,tdm3,tdmastro,tdm4,tdm5,tdm6,bastard}. The invention of Graphene had boosted the interest of applying effective mass in quantum field theory, in particular for the Dirac equation. In this article we shall use one dimensional Dirac equation with mass and fermi velocity which are constant except for a finite jump at a point in real axis. 
 \begin{eqnarray}\label{velocity}
 v(x)=\begin{cases}
 v_l, & \text{if} \;\;\; x < 0 \\
 v_r, & \text{if} \;\;\; x > 0,
 \end{cases}
 \end{eqnarray}
 And
 \begin{eqnarray}\label{mass}
 m(x)=\begin{cases}
 m_l, & \text{if} \;\;\; x < 0 \\
 m_r, & \text{if} \;\;\; x > 0.
 \end{cases}
 \end{eqnarray} 
 The Dirac Hamiltonian for fermions inside graphene in $1+1$ dimension is given by
 \begin{equation}\label{hamiltonian}
 \hat{H}=\frac{\hbar}{i}v(x)\sigma_x\frac{d}{dx}+ m(x)v(x)^2\sigma_z .
 \end{equation}
 Where $\sigma_x,\sigma_y$ and $\sigma_z$ satisfy the Clifford algebra. We shall use the following representation given by Pauli matrices
 
 \begin{eqnarray}
 \sigma_x=\left(\begin{array}{cc}
 0 & 1\\
 1 & 0
 \end{array}\right), \;\;\;
 \sigma_y=\left(\begin{array}{cc}
 0 & -i\\
 i & 0
 \end{array}\right),\;\;\;
 \sigma_z=\left(\begin{array}{cc}
 1 & 0\\
 0 & -1
 \end{array}\right).
 \end{eqnarray}
 From now on, we shall use $\hbar=1$.\\
 One may be tempted to write down the eigenfunctions of the Hamiltonian as 
 follows
 \begin{eqnarray}\label{eigenfunction}
 \psi(x)=\left\{\begin{array}{ccc}
 \psi_l = & A \left(\begin{array}{c}
 1\\
 \sqrt{\frac{E_l-m_lv_l^2}{E_l+m_lv_l^2}} 
 \end{array}\right)e^{\kappa_l x}, & \mbox{for}\; x<0 .\\
 \psi_r = & B \left(\begin{array}{c}
 1\\
 -\sqrt{\frac{E_r-m_rv_r^2}{E_r+m_rv_r^2}} 
 \end{array}\right)e^{-\kappa_r x} ,&\mbox{for}\; x>0.
 \end{array} \right.\\
 \mbox{With,} \;\;\kappa_l=m_l^2v_l^2-\frac{E_l^2}{v_l^2} ,\mbox{and,}\;\; \kappa_r=m_r^2v_r^2-\frac{E_r^2}{v_r^2} .
 \end{eqnarray}
 But, this would create a self-contradiction in the matching condition of eigenvalue equation of the Hamiltonian. Problem lies in the matching condition at around $x=0$. A proper extension of $\hat{H}$ is required so that it becomes self-adjoint. One may show that infinitely many extensions are possible. But, we want to confine our interest in the situation where particle confinement is possible inside the graphene. To do so, at the origin, one can include a pure scalar point interaction potential of the form 
 \begin{equation}
 \mathcal{U}(x)=\sigma_z g_s u(x).
 \end{equation}
 Where, $u(x)$ is any function which has pick at $x=0$ and at satisfy $\int_{-\infty}^{\infty} u(x)dx=1$. For example one can take a Dirac delta type distribution $(\delta(x))$. From now on we shall write the strength of the function as $g_s=a$. One can show that, under this condition, one of the  matching conditions at $x=0$ may be given by
 \begin{eqnarray}\label{matchingcondition}
 \left(\begin{array}{c}
 \phi_1(0+) \\
 \phi_2(0+)
 \end{array}\right) =
  \left(\begin{array}{cc}
 \left(\frac{1+m_r^2v_F^4}{1+m_l^2v_F^4}\right)^{\frac{1}{4}} \cosh\left(\frac{a}{v_F}\right) & i\sinh\left(\frac{a}{v_F}\right)\\
  -i\sinh\left(\frac{a}{v_F}\right)& \left(\frac{1+m_l^2v_F^4}{1+m_r^2v_F^4}\right)^{\frac{1}{4}}\cosh\left(\frac{a}{v_F}\right) 
 \end{array}\right) =
 \left(\begin{array}{c}
 \phi_1(0-) \\
 \phi_2(0-)
 \end{array}\right).
 \end{eqnarray}
 To be more specific, let us consider that the incoming wave from the left side of origin is monochromatic. Then, one can write down the eigenfunction of \eqref{hamiltonian} as
 \begin{eqnarray}\label{monochromatic}
 \psi(x)=\left\{\begin{array}{cc}\left(\begin{array}{c}
 1 \\
 \alpha_l
 \end{array}\right)e^{ik_lx}+
 r_1\left(\begin{array}{c}
 1 \\
 -\alpha_l
 \end{array}\right)e^{-ik_lx}, & x<0.\\
 t_1\left(\begin{array}{c}
 1 \\
 \alpha_r
 \end{array}\right)e^{ik_rx}, & x>0.
 \end{array}
 \right.
 \end{eqnarray}
 Where,
 \begin{eqnarray}\label{alphalalphar}
 \alpha_l= \sqrt{\frac{E-m_lv_F^2}{E+m_lv_F^2}},\\
 \alpha_r= \sqrt{\frac{E-m_rv_F^2}{E+m_rv_F^2}} .\nonumber
 \end{eqnarray}
 Here, $r_1$ and $t_1$ are reflection and transmission amplitude respectively. They can be determined in terms of 'a' (strength of scalar PIP), $m_l,\;m_r$ and $v_F$ by using \eqref{matchingcondition}. 
 \section{Information Entropy}
 In order to calculate the information entropy, we need the probability density which reads in $\{\vert x \rangle\}$ space as follows.
 \begin{eqnarray}\label{probabilitydensity}
 \rho(x)= \left\{
 \begin{array}{ccc}
 \alpha+\beta\cos(2k_lx), &\mbox{for}& x<0. \\
 t_1^2\left(1+\alpha_r^2\right), &\mbox{for}& x>0. 
 \end{array}
 \right.\\
 \mbox{with,}\;\; \alpha= \left(1+r_1^2\right)\left(1+\alpha_l^2\right), \; \beta=2r_1\left(1-\alpha_l^2\right).
 \end{eqnarray}
 One can note that 
 \begin{eqnarray}
 \because r_1+\frac{1}{r_1}\ge 2, \;\forall r \in \mathbb{R}^+ , \nonumber \\ 
 \mbox{and}, \; E> m_lv_F^2 ,\nonumber\\
 \therefore \frac{\alpha}{\beta}=\frac{E}{2m_lv_F^2}\left(r_1+\frac{1}{r_1}\right) >1 .
 \end{eqnarray}
 Therefore, $\rho(x)$ is positive for every value of $x$. Moreover, $\rho(x)$ is nonzero. Because, if  it becomes zero at any point in real line $(\mathbb{R})$, then $\cos 2k_lx=-\frac{\alpha}{\beta}$ at that point. In that case
 \begin{eqnarray}\label{inequalityforbranch}
 -1\le \frac{\alpha}{\beta}\le  1 .\nonumber \\
 \therefore -\frac{m_lv_F^2}{E}\le \frac{1}{2}\left(r_1 + \frac{1}{r_1}\right) \le \frac{m_lv_F^2}{E} <1 .\nonumber \\
 \therefore r_1 +\frac{1}{r_1}< 2 .
 \end{eqnarray}
 But, \eqref{inequalityforbranch} is a contradiction. 
 Therefore, $s_x=\rho(x)\log\rho(x)$ is nonsingular on entire $\mathbb{R}$.\\
 Because of the periodicity of $s_x$ and $\rho(x)$ for $x<0$ region and as well as on $x>0$ region separately, we split $\mathbb{R}^{-}$ in $mod(\frac{\pi}{k_l})$ and assuming infinity as a  multiple of $\frac{\pi}{k_l}$ and due to $s_x(-x)=s_x(x)$, we can write
 \begin{equation}
 \int_{-\infty}^{0} s_x dx =\int_0^\infty s_x dx= \lim_{N\to \infty} N\int_0^{\frac{\pi}{k_l}} s_xdx.
 \end{equation} 
 And similarly for $x>0$ region.\\
 Then the Shannon information entropy is given by
 \begin{eqnarray}
 S_x=-<s_x>
 =- \frac{\int_{-\infty}^{\infty}\rho(x) \log \rho(x) dx}{\int_{-\infty}^{\infty}\rho(x)dx} \nonumber \\
 =- \frac{1}{\vert\vert \psi(x)\vert\vert^2}\lim_{N\to \infty}N\int_{0}^{\frac{\pi}{k_l}}\{\left(\alpha+\beta\cos 2k_lx\right) \log\left(\alpha+ \beta\cos 2k_lx\right)   + \nonumber \\ t_1^2\left(1+\alpha_r^2\right)\log \left(t_1^2\left(1+\alpha_r^2\right)\right)\}dx .
 \end{eqnarray}
 Where,
 \begin{eqnarray}
 \vert\vert \psi(x)\vert\vert^2 =  \lim_{N\to \infty}\left[N\int_{0}^{\frac{\pi}{k_l}}\left(\alpha+\beta\cos 2k_lx\right)dx + N\int_{0}^{\frac{\pi}{k_l}}t_1^2\left(1+\alpha_r^2\right)dx\right] \nonumber \\
 = \lim_{N\to\infty} N\frac{\pi}{k_l}\left[\alpha+t_1^2\left(1+\alpha_r^2\right)\right].
 \end{eqnarray}
 We can further identify that
 \begin{eqnarray}
 I=\int\limits_{0}^{2\pi}\left(\alpha+\beta\cos x\right)\log\left(\alpha+\beta\cos x\right)dx \nonumber\\
 =I_{1+}+I_{1-}+I_{2+}+I_{2-}.
 \end{eqnarray} 
 Where
 \begin{eqnarray}
 I_{1\pm}=\int\limits_{0}^{\frac{\pi}{2}} \left(\alpha\pm \beta\cos x\right)\log\left(\alpha\pm\beta\cos x\right)dx .\\
 I_{2\pm}=\int\limits_{0}^{\frac{\pi}{2}} \left(\alpha\pm \beta\sin x\right)\log\left(\alpha\pm\beta\sin x\right)dx.
 \end{eqnarray}
 That means
 \begin{eqnarray}
 \because 0\le \cos x , \sin x \le 1 , \;\mbox{for}\; 0\le x\le\frac{\pi}{2} .\nonumber \\
 \frac{\pi}{2}\alpha\log\alpha \le I_{1+}, I_{2+} \le \frac{\pi}{2}\left(\alpha+\beta\right)\log\left(\alpha+\beta\right).\\
 \frac{\pi}{2}\left(\alpha-\beta\right)\log\left(\alpha-\beta\right) \le I_{1-}, I_{2-} \le \frac{\pi}{2}\alpha\log\alpha .
 \end{eqnarray}
 Therefore, we can conclude that
 \begin{eqnarray}
 S_x\ge -\frac{\frac{\alpha}{2}\log\alpha +\frac{1}{2}\left(\alpha+\beta\right)\log\left(\alpha+\beta\right)+ t_1^2\left(1+\alpha_r^2\right)\log\left(t_1^2\left(1+\alpha_r^2\right)\right)}{\alpha+t_1^2\left(1+\alpha_r^2\right)}.
 \end{eqnarray}
 Now, let us fix the normalization of FT s.t. the F.T of $\psi(x)$ is defined by
 \begin{equation}\label{FTdefinition}
 \tilde{\phi}(p)=\frac{1}{\sqrt{2\pi}}\int\limits_{-\infty}^{\infty}\psi(x)e^{-ipx}dx.
 \end{equation}
  Using ~\eqref{monochromatic} in ~\eqref{FTdefinition} one gets
  \begin{eqnarray}
  \tilde{\Phi}(p)=\frac{i}{\sqrt{2\pi}} \left( 
  \begin{array}{c}
  \frac{t_1}{k_r +p}+\frac{(1+r_1)p-(1-r_1)k_l}{k_l^2-p^2} \\
  \frac{t_1\alpha_r}{k_r +p}+\frac{\alpha_l(1-r_1)p-\alpha_l(1+r_1)k_l}{k_l^2-p^2} \end{array}\right).
  \end{eqnarray}
 Now the probability density in momentum space reads
 \begin{eqnarray}\label{densityinmomentumspace}
 \tilde{\rho}(p)= \tilde{\Phi}(p)^\dag \tilde{\Phi}(p) \\
 = \frac{\delta_0}{2\pi}\frac{(p-p_1) (p-p_2) (p-p_3) (p-p_4)}{(p+k_r)^2 (p^2-k_l^2)^2}.
 \end{eqnarray}
 The entropy in momentum space is given by
 \begin{equation}\label{entropymomentumdefinition}
 S_p=- \frac{\int_{-\infty}^{+\infty} \tilde{\rho}(p)\log \tilde{\rho}(p)dp}{\int_{-\infty}^{+\infty}\tilde{\rho}(p)dp}.
 \end{equation}
 which is nonzero positive quantity. That means 
 \begin{eqnarray}
 S_x+S_p \ge -\frac{\frac{\alpha}{2}\log\alpha +\frac{1}{2}\left(\alpha+\beta\right)\log\left(\alpha+\beta\right)+ t_1^2\left(1+\alpha_r^2\right)\log\left(t_1^2\left(1+\alpha_r^2\right)\right)}{\alpha+t_1^2\left(1+\alpha_r^2\right)}.
 \end{eqnarray}
 Therefore, an uncertainty relation in terms of SE exists for the system under consideration.
 \section{Conclusions}
 We have seen that the entropy uncertainty relation (UR) holds for the case of position-dependent mass. In particular, for the case of particle confinement in the graphene, where the effective mass will have a jump discontinuity at origin will exhibit a kind of UR. Our study is limited in the sense that the exact result in momentum space is not obtained. However, by showing the existence of a lower bound for uncertainty relation for the  Dirac System with a mass jump, our aim regarding this article is fulfilled. Further study for the systems with mass jump under some external potential term will be interesting for the future study.
 \section{Acknowledgement} 
 PP is grateful to J P Saha for warm hospitality and stimulating discussions at the Department of Physics, University of Kalyani.

 \end{document}